\documentclass[twocolumn,showpacs,preprintnumbers,superscriptaddress,amsmath,amssymb]{revtex4-2}
\usepackage{amssymb}		
\usepackage{amsmath}
\usepackage{graphicx}
\usepackage[normalem]{ulem}
\usepackage{multirow}
\usepackage{appendix}
\usepackage[usenames]{color}
\usepackage{bm}
\usepackage{CJK}
\usepackage[colorlinks,linkcolor=blue,urlcolor=blue,citecolor=blue]{hyperref}
\usepackage{booktabs} 	
\usepackage{leftidx}

\usepackage[all]{hypcap} 
\makeatletter

\newcommand{\Rmnum}[1]{\expandafter\@slowromancap\romannumeral #1@}
\makeatother

\newcommand\redsout{\bgroup\markoverwith{\textcolor{red}{\rule[0.5ex]{2pt}{0.4pt}}}\ULon}

\usepackage{tikz,xcolor,hyperref}

\definecolor{lime}{HTML}{A6CE39}
\DeclareRobustCommand{\orcidicon}{
	\begin{tikzpicture}
	\draw[lime, fill=lime] (0,0) 
	circle [radius=0.16] 
	node[white] {{\fontfamily{qag}\selectfont \tiny ID}};
	\draw[white, fill=white] (-0.0625,0.095) 
	circle [radius=0.007];
	\end{tikzpicture}
	\hspace{-2mm}
}
\foreach \x in {A, ..., Z}{%
	\expandafter\xdef\csname orcid\x\endcsname{\noexpand\href{https://orcid.org/\csname orcidauthor\x\endcsname}{\noexpand\orcidicon}}
}

\begin{document}
	\begin{CJK*} {UTF8} {gbsn}

\title{System dependence of away-side broadening and  $\alpha$-clustering light nuclei structure effect in dihadron azimuthal correlations}

\author{Yuan-Zhe Wang (王远哲)}
\affiliation{Key Laboratory of Nuclear Physics and Ion-beam Application (MOE), Institute of Modern Physics, Fudan University, Shanghai 200433, China}

\author{Song Zhang (张松)\orcidB{}}\email{song\_zhang@fudan.edu.cn}

\affiliation{Key Laboratory of Nuclear Physics and Ion-beam Application (MOE), Institute of Modern Physics, Fudan University, Shanghai 200433, China}
\affiliation{Shanghai Research Center for Theoretical Nuclear Physics， NSFC and Fudan University, Shanghai 200438, China}

\author{Yu-Gang Ma (马余刚)\orcidC{}}
\email{mayugang@fudan.edu.cn}
\affiliation{Key Laboratory of Nuclear Physics and Ion-beam Application (MOE), Institute of Modern Physics, Fudan University, Shanghai 200433, China}
\affiliation{Shanghai Research Center for Theoretical Nuclear Physics， NSFC and Fudan University, Shanghai 200438, China}

\begin{widetext}
\end{widetext}

\begin{abstract}
 A collision system scan involving $\alpha$-clustered $^{12}$C and $^{16}$O is studied by using a multiphase transport model for  central collisions at $\sqrt{s_{NN}} = 6.37$ TeV. Background subtracted away-side dihadron  azimuthal correlation is performed via the zero yield at minimum (ZYAM) method from raw signals, and the quantitative parameters, such as RMS width and Kurtosis, seem nicely follow the $A^{-1/3}$ law of the system size if the nucleus has the normal Woods-Saxon nucleon distribution. However, for	 $\alpha$-clustering light nuclei, specifically for $^{12}$C and $^{16}$O, the RMS width and Kurtosis of away-side azimuthal correlation are deviated from the baseline of $A^{-1/3}$ law.  In addition, the momentum dependence of away-side  broadening  
 parameters is also presented. The results show  that there is a distinction in 
	away-side broadening parameters of dihadron correlation function  between the  Woods-Saxon distribution and the $\alpha$-clustered structures, which  sheds light on that the collision system scan for dihadron azimuthal correlation as a potential probe to distinguish $\alpha$-clustered nuclei. 
\end{abstract}

\maketitle

\section{Introduction}
Ultra-relativistic heavy-ion collisions provide an effective environment for the study of strongly interacting partons. In the collisions, participant nucleons melt into deconfined quarks and gluons, leading to formation of a hot dense matter, which is so-called quark-gluon plasma (QGP). Energetic partons created in hard scattering process are believed to lose considerable energy when they traverse the dense medium. For particles spread from the primary vertex with different azimuthal direction, they will travel different distance and nuclear medium thickness inside the collision region.  The energy loss  known as jet quenching~\cite{gyulassyHIJINGMonteCarlo1994, wangJetQuenchingAzimuthal2001, armestoMeasuringCollectiveFlow2004, ZHANG201176} is dominated by path length and nuclear medium thickness~\cite{wangJetQuenchingAzimuthal2001}. As a result, some consequential observables with azimuthal anisotropy emerge, for instance, the strong suppression on the yields of away-side high $p_T$ particles~\cite{adlerDisappearanceBackToBackHigh2003, adamsAzimuthalAnisotropyCorrelations2004}. Even though the detailed mechanism of jet quenching is still under debate, such a phenomenon gives clues on the collective motion of QGP and is expected to reflect properties of the pre-collision system. However, it is difficult to detect the jet fragmentation process directly, whereas dihadron azimuthal correlation distribution is a widely used method to reconstruct the picture of jet signals~\cite{ULERY2006581, ajitanandDecompositionHarmonicJet2005, maDihadronAzimuthalCorrelation2006}, and the information of the early stage can be learned by studying jet quenching phenomenon as well~\cite{wangJetQuenchingAzimuthal2001, agakishievMeasurementsDihadronCorrelations2021,  agakishievEventplanedependentDihadronCorrelations2014, adareMeasurementTwoparticleCorrelations2019,WangPRL,WangPRL2,WangPRL3}. 

	Over the past decades, there have been numerous experiments that carried out dihadron azimuthal correlation measurements at various center-of-mass energies as well as different collision systems~\cite{adler_dense-medium_2006, aggarwalAzimuthalDihadronCorrelations2010, adamczykDihadronCorrelationsIdentified2015, ATLAS6-1, CMS6-2, aadObservationAssociatedNearSide2013, aadObservationLongRangeElliptic2016, agakishievMeasurementsDihadronCorrelations2021,STAR_CPC}. It was found that small collision systems such as p + p and p + A in high-multiplicity present resembling properties with large A + A collisions~\cite{alicecollaborationEnhancedProductionMultistrange2017}. Due to minor size and short lifetime, small-size A + A systems may undergo a modified dynamical process and are expected to shed light on initial fluctuation effects on momentum distribution in the final stage~\cite{correlation5, agakishievEventplanedependentDihadronCorrelations2014, adareMeasurementTwoparticleCorrelations2019, nieInfluenceInitialstateMomentum2019, zhangCollisionSystemSize2020}.
	
	Concerning light nuclei systems, some interesting structure phenomena provoke ones interest, eg. for $\alpha$-clustering structure.  One of the most famous work was the Hoyle state of $^{12}$C~\cite{hoyle}. 	 So far, $\alpha$ cluster model has attracted much attention~\cite{freerMicroscopicClusteringLight2018} since it was postulated by Gamow~\cite{Gamow1931alpha}. Both theoretical and experimental efforts provided crucial evidence for the formation of $\alpha$-particle condensates in light self-conjugate nuclei~\cite{beckClustersNuclei2010, tohsakiAlphaClusterCondensation2001,YeYL2018,Zhou1}.
	In particular, for $^{12}$C and $^{16}$O, $\alpha$-clustering behavior is rather common even in ground state.
	Compared with classical Woods-Saxon distribution, nuclei with $\alpha$-clustering structure could emerge a deformed intrinsic configuration. The ground state of $\alpha$-clustered $^{12}$C is suggested to show a triangular-like arrangement, while $\alpha$-clustered $^{16}$O has a tetrahedron structure ~\cite{vonoertzenNuclearClustersNuclear2006, heGiantDipoleResonance2014, arriolaLowEnergyNuclear2015, rybczynskiSignaturesClusteringUltrarelativistic2018,Shi2021}. Specific spatial nucleon distribution incites us to search for different physical features between clustered nuclei and non-clustered cases. For example, our previous works demonstrated the effects of $\alpha$-clustering structure on giant dipole resonance~\cite{heGiantDipoleResonance2014,Huang3} and photonuclear reactions~\cite{huangPhotonuclearReactionProbe2017, huangTwoprotonMomentumCorrelation2020,Huang4}.

	It has been proposed that  initial $\alpha$-clustering nuclear structure could be distinguished via high energy heavy-ion collisions~\cite{broniowskiSignaturesClusteringLight2014}. This methodology originates from hydrodynamic calculations, which demonstrate that the asymmetry of the initial coordinate space shall turn into the momentum space anisotropy during the evolution of deformed fireball whose shape relates to original nucleon distribution~\cite{teaney_unusual_1999, galeHYDRODYNAMICMODELINGHEAVYION2013,Yan2020,CPL21}. Many work focus on collective flow as suggested in Ref.~\cite{broniowskiSignaturesClusteringLight2014,He2}, revealing evidential distinction between uniform and clustered nuclei~\cite{bozek__2014, zhangNuclearClusterStructure2017, rybczynskiSignaturesClusteringUltrarelativistic2018,PRC21,He1}. 
	
	To investigate the influence on dihadron azimuthal correlations from the intrinsic geometry in $\alpha$-clustered nuclei, the system scan project was proposed in our previous work with observables of collective flow ratio and forward-backward multiplicities correlations~\cite{liSignaturesClustering162020,Li2021}  as a measurable method. In the present work,  a system scan was employed by using a multi-phase transport (AMPT) model~\cite{linMultiphaseTransportModel2005} for most central symmetric collision systems from $^{10}$B + $^{10}$B to $^{197}$Au + $^{197}$Au at $\sqrt{s_{NN}} = 6.37$ TeV. The dihadron azimuthal correlation functions were calculated and the background with respective to different order event planes were reconstructed. Based on the correlation functions, the system size dependence of the root-mean-square width and kurtosis in the away-side region is discussed. These results shed light on the future system scan project at  LHC~\cite{RHIC-smallsys-propose,LHC-samllsys-propose}, from which  the away-side broadening structure was proposed as a probe to distinguish the $\alpha$-clustering structure in light nuclei, such as $^{12}$C and $^{16}$O.
 
	The paper is arranged as follows: In Sec.~\ref{sec:Methods}, we introduce the AMPT model and the procedures for evaluating dihadron azimuthal correlation. The correlation results and discussion are placed in Sec.~\ref{sec:Results}. Several relevant cumulants focusing on the away-side signals are presented simultaneously. Finally we give a summary in Sec.~\ref{sec:Summary}.

\section{\label{sec:Methods}AMPT Model and Analysis Methods}

	In this work, we use a multiphase transport model (AMPT)~\cite{linMultiphaseTransportModel2005,AMPT2021} to perform calculations for collision systems with different sizes in central collisions. The AMPT model consists of four main physics stages, which describe the four main processes in heavy-ion collisions. The initial conditions including the generation of partons are simulated by the Heavy Ion Jet Interaction Generator (HIJING) model~\cite{HIJING-1,HIJING-2}, the partonic interaction is modeled with Zhang’s parton cascade (ZPC) model~\cite{ZPCModel} with only 2 to 2 elastic parton process at present, the hadronization process is carried through a quark coalescence model, and finally the hadron scattering is described by A Relativistic Transport (ART) model~\cite{ARTModel}. 
	 There are two AMPT versions: one for a string melting mechanism, in which a partonic phase is generated from excited strings in the HIJING model, and a simple quark coalescence model is used to combine the partons into hadrons; the another for the default AMPT version which only undergoes a pure hadron gas phase.  
	 AMPT model has been widely used in heavy-ion collision at RHIC and LHC~\cite{linMultiphaseTransportModel2005,AMPT2021,AMPTGLM2016,Wang1,Wang2,Tang} and the detailed introduction can be found in Refs.~ \cite{linMultiphaseTransportModel2005,AMPT2021}. In the present work, we essentially use the string-melting version AMPT which is more suitable to treat parton dominated interaction process in early stage collision dynamics at  ultra-relativistic energy, but the default AMPT model is also utilized  to investigate the effect from partonic interactions. The presented figures are the results based on the string-melting AMPT version except it is particularly mentioned using the default AMPT version.

	For initial nucleon distributions of different systems, namely $^{10}$B, $^{12}$C, $^{16}$O, $^{40}$Ca, $^{96}$Zr, and  $^{197}$Au systems, the Woods-Saxon (WS) distribution of nucleons are introduced from the HIJING model.  However, considering the $\alpha$-clustering structure of light nuclei,  $^{12}$C and $^{16}$O are also set up as triangular and tetrahedral $\alpha$ structures in initial state of the AMPT model in order to compare the results for their WS nucleon distributions. 
	The original parameters of $\alpha$-clustered $^{12}$C and $^{16}$O nuclei  were calculated  by the EQMD model with effective Pauli potential ~\cite{maruyamaExtensionQuantumMolecular1996}. In order to match the experimental data~\cite{angeliTableExperimentalNuclear2013}, vertexes where $\alpha$ cluster lays in the triangular-liked $^{12}$C are placed 3.10 fm away from each other~\cite{heGiantDipoleResonance2014, zhangNuclearClusterStructure2017}. For the tetrahedral pattern of $^{16}$O, we assign the side length as 3.42 fm~\cite{liSignaturesClustering162020,Li2021,Li3}. Meanwhile, $\alpha$ clusters which contain four nucleons inside are set in accord with the Woods-Saxon distribution. For each clustered nucleus we set a random orientation. The centrality of event is determined by the multiplicity within $0.2\ < p_T < 5\ \mathrm{GeV/c}$ and $|\eta| < 1$.

	At the early stage of the relativistic heavy-ion collisions, jets are always produced back-to-back on the transverse plane in the hard scattering processes~\cite{connorsJetMeasurementsHeavy2018}.  The particle in the jets will loss part of its energy while passing through the QGP matter created in the collisions.  
	The energy and momentum will be redistributed via jet quenching and then the correlation emerges among the particles. Dihadron azimuthal correlation  is constructed by paired hadrons, where a high $p_T$ hadron denoted as a trigger particle stemming from hard scattering processes and a lower $p_T$ hadron as an associated particle which may inherit energy or momentum from the trigger particle. The dihadron azimuthal correlation is defined as $\Delta\phi = \phi_{\mathrm{asso}} -\phi_{\mathrm{trig}} $ distribution, 
\begin{equation}
C(\Delta\phi) = \frac{1}{N_{trig}}\frac{dN}{d\Delta\phi},
\label{eq:DPhiCorr}
\end{equation}
where $N_{trig}$ is the number of trigger particles, $\phi_{\mathrm{asso}}$ and $\phi_{\mathrm{trig}}$ are the azimuthal angle of the associated particle and the trigger particle, respectively.

\begin{figure}[thb]
	\centering
		\includegraphics[angle=0,scale=0.45]
		{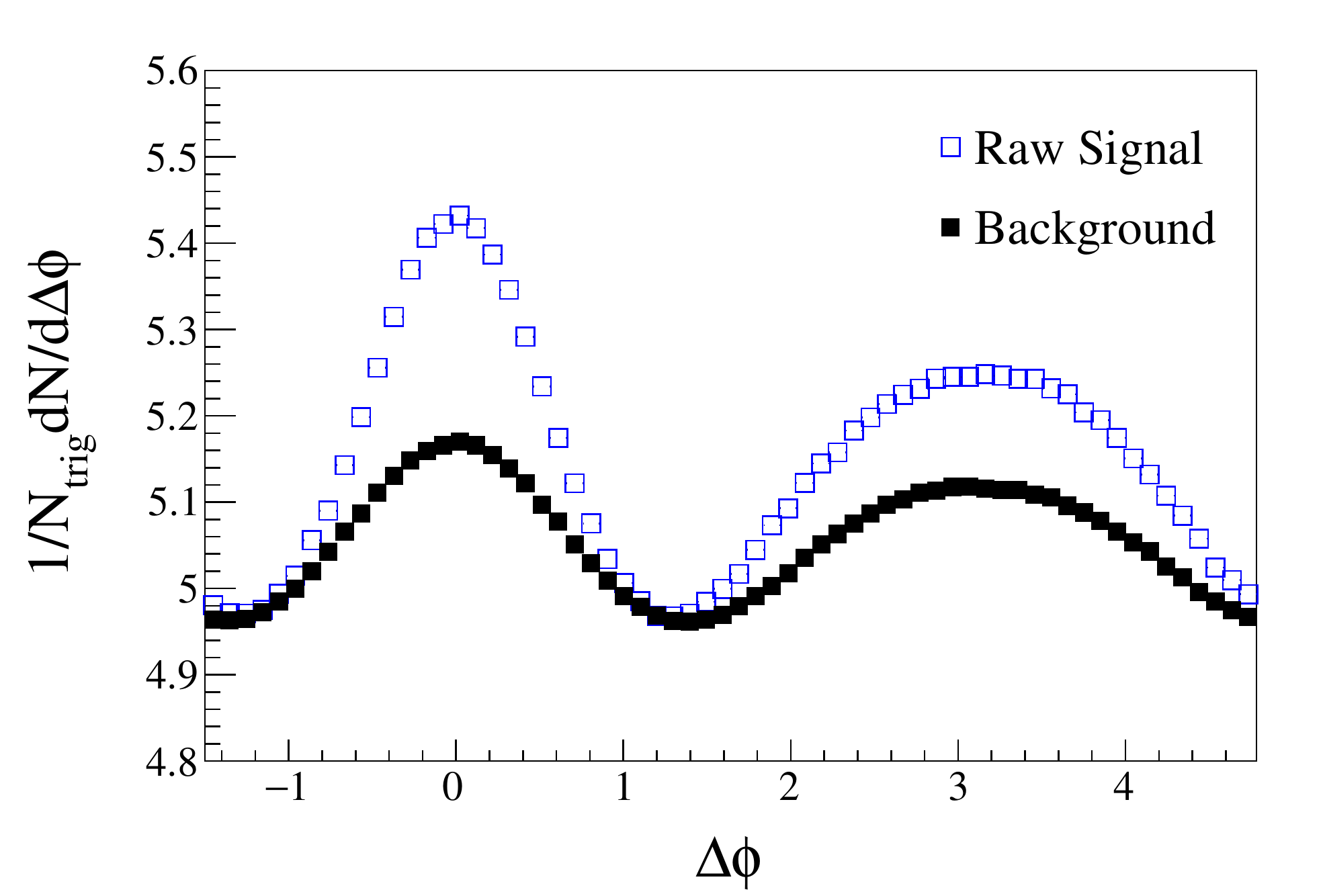}
		\caption{ dihadron azimuthal correlation signal (open circles) and background (solid squares) reconstructed by mix-events method for tetrahedron clustered $^{16}$O + $^{16}$O collisions at $\sqrt{s_{NN}} = 6.37$ TeV in 0--2\% centrality, both are divided by total numbers of trigger particles in all events. The background of per-trigger yields $\Delta\phi$ distribution is reconstructed by mixing events with the similar second and third order event planes and normalized with ZYAM method.}
		\label{fig:SignalBack}
\end{figure}

The most significant background in the dihadron azimuthal correlations comes from the collective flow, which can be expanded as~\cite{adamsDistributionsChargedHadrons2005},
	\begin{equation}
		f(\Delta\phi)  \propto \left[1+\sum_{n=1}^{\infty}2\langle v^a_n v^b_n\rangle \cos(n\Delta\phi)\right],\label{eq2}
	\end{equation}
	where $v^a_n$ and $v^b_n$ are $n$-th order collective flow coefficients of trigger and associated particles. To estimate this background $\Delta\phi$ distribution, a mixed-event method was used~\cite{maLongitudinalBroadeningNearside2008, adamsDistributionsChargedHadrons2005} as following: the trigger particle and associated particle are from different events which have similar physical properties, such as the second and third order event plane angles which will be introduced later. Note that all collision systems we used in this analysis are central collisions within a small centrality bin. The effect of multiplicity or the centrality, therefore, is negligible in the mixed-event method. In practice, $dN_{mixed}/d\Delta\phi$ and $dN_{same}/d\Delta\phi$, denote the $\Delta\phi$ distributions from mixed events and same events, respectively, and then the real signal of correlation function~(\ref{eq:DPhiCorr}) is obtained via $C(\Delta\phi) = 1/N_{trig}(dN_{same}/d\Delta\phi - B_0*dN_{mixed}/d\Delta\phi$), where $B_0$ is a normalized factor by a so-called ZYAM method~\cite{ajitanandDecompositionHarmonicJet2005} (zero yield at minimum).
	To illustrate the ZYAM method, figure~\ref{fig:SignalBack} shows the correlation function constructed in the same event (raw signal) and mixed events (background), respectively, for tetrahedron clustered $^{16}$O + $^{16}$O collisions at $\sqrt{s_{NN}}$ = 6.37 TeV. The normalized factor $B_0$ can be tuned to get zero yield in the correlation function $C(\Delta\phi)$ in a region of $0.8<\Delta\phi<1.2$.

In order to construct the background, a widely-used approach to estimate the reaction plane called event plane is applied here. The event plane is gained by the orientation of the flow vector ~\cite{poskanzerMethodsAnalyzingAnisotropic1998}. The $n$-th order event plane angle is defined as,
	\begin{equation}
		\Psi_{n}^{\mathrm{EP}}=\tan ^{-1} \left(\frac{\sum_{i} \sin \left(n \phi_{i}\right)}{\sum_{i} \cos \left(n \phi_{i}\right)}\right),
	\end{equation}
	where $\phi_{i}$ is the azimuthal angle of particle $i$, the sum runs over all emitted particles in the event. Noting that the event plane angle calculated above are only used in the mixing event procedure for estimating the background.

\section{\label{sec:Results}Results and Discussion}

	The dihadron azimuthal correlations are calculated in central A + A collisions with kinetic windows, i.e. rapidity in $|y|<1$ and transverse momentum $2<p_T<6~\mathrm{GeV}/c$ for the trigger particle and $0.2<p_T<2~\mathrm{GeV}/c$ for associated particles. And the charged hadrons $\pi^\pm$, $K^\pm$, $p$ and $\bar{p}$ are selected for this study.

\begin{figure}[hb]
	\centering
		\includegraphics[angle=0,scale=0.45]
		{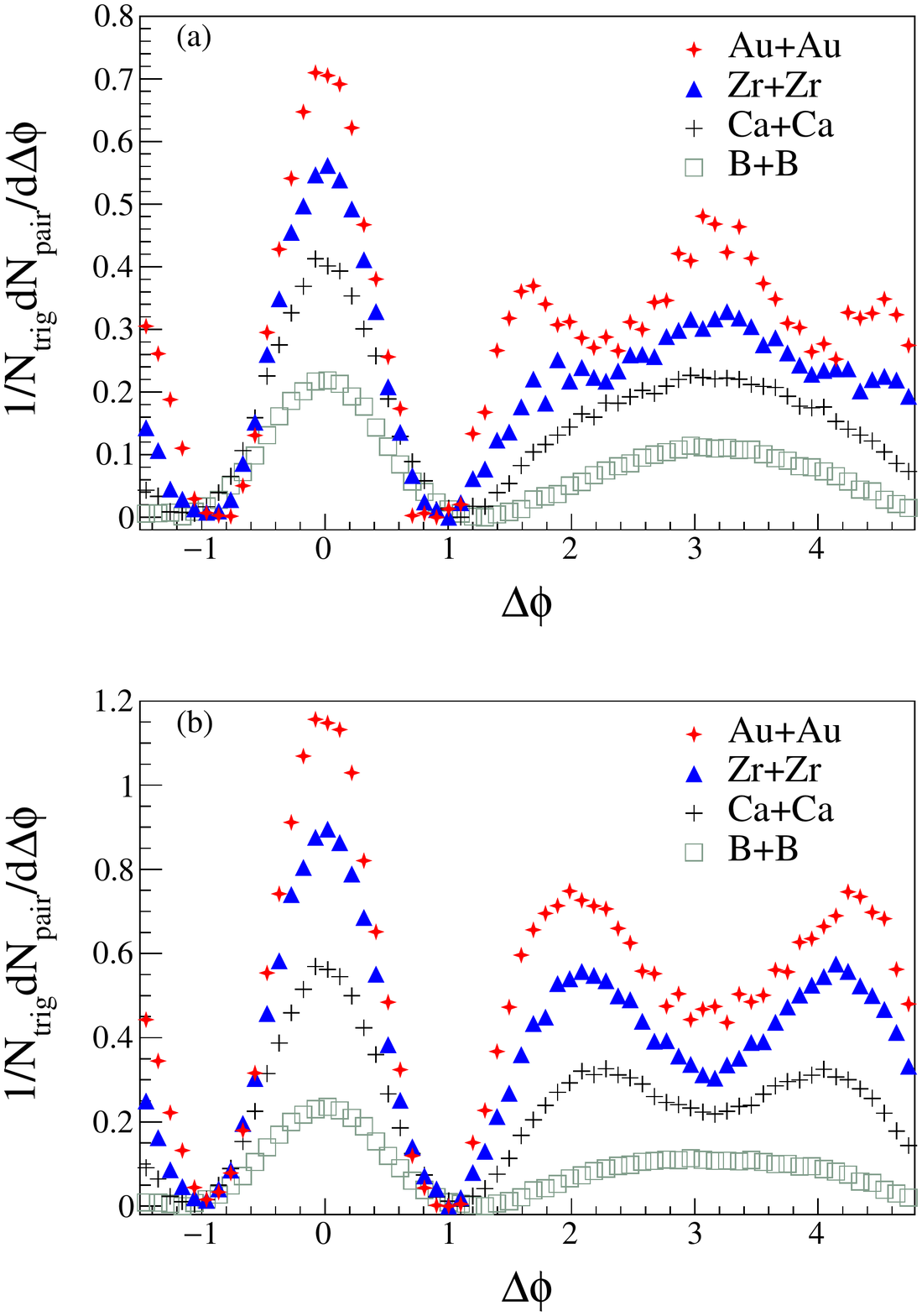}
		\caption{ Subtracted dihadron azimuthal correlation function in different collision systems of nuclei with the Woods-Saxon distributions in 0--2\% centrality at $\sqrt{s_{NN}} = 6.37$ TeV. Green squares for $^{10}$B + $^{10}$B, black plus for $^{40}$Ca + $^{40}$Ca, blue triangles for $^{96}$Zr + $^{96}$Zr, red double-diamond for $^{197}$Au + $^{197}$Au. Panel (a) deduct both the second and the third order background, while panel (b) just subtracts the second order background.}\label{fig:CorrelationAll}
\end{figure}

\begin{figure*}[thb]
	\centering
		\includegraphics[angle=0,scale=0.80]
		{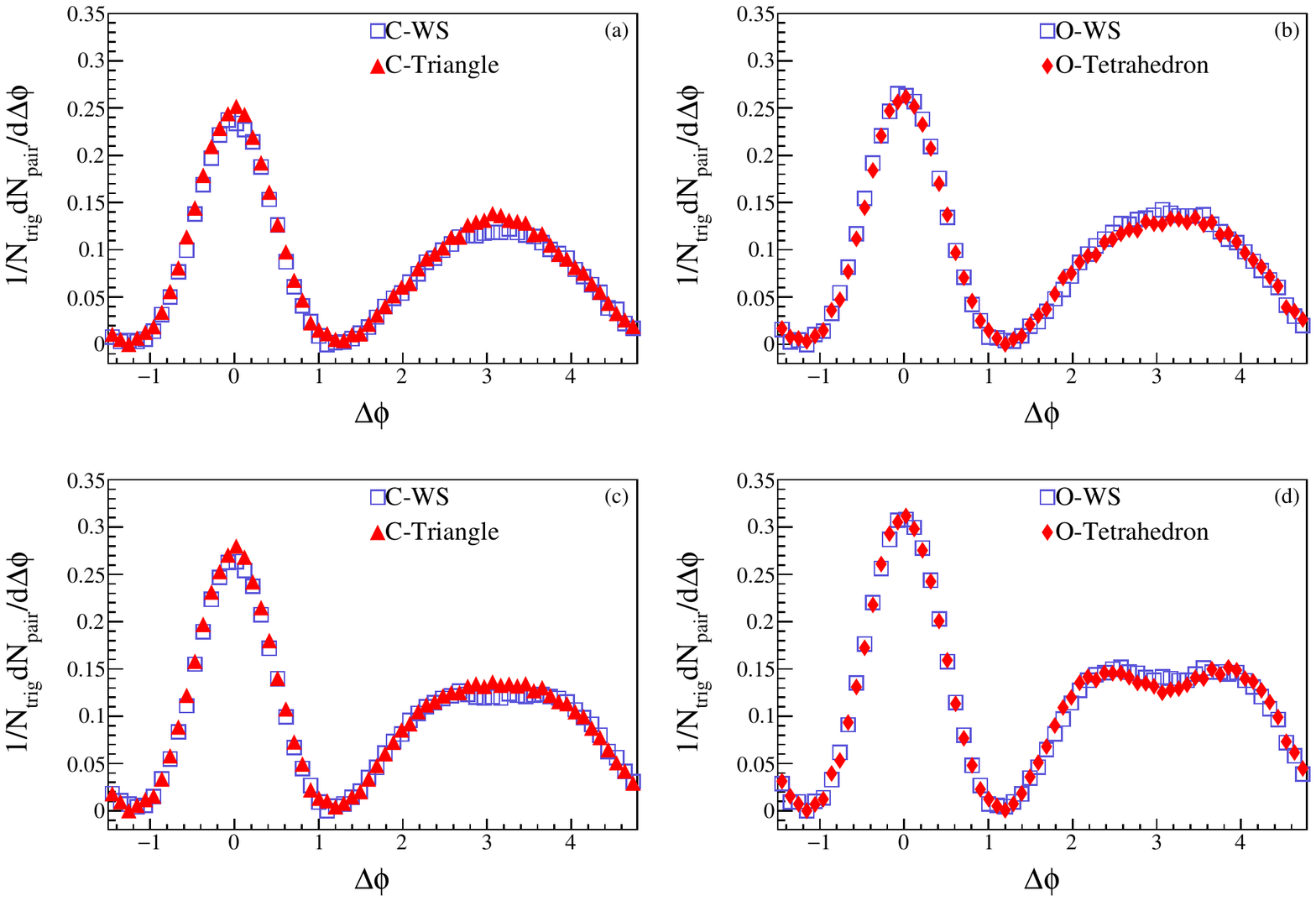}
		\caption{ Subtracted dihadron azimuthal correlation function in $^{12}$C + $^{12}$C (left panel) and $^{16}$O + $^{16}$O (right panel) collision systems at $\sqrt{s_{NN}} = 6.37$ TeV in 0--2\% collisions. Open squares represent the Woods-Saxon distribution for nucleus in both systems, red triangles for triangular 3$\alpha$-clustered $^{12}$C, and red diamonds for the $^{16}$O with tetrahedron 4$\alpha$ arrangement. Panel (a) and (b) deduct both the second and the third order background, (c) and (d) just subtract the second order background.}\label{fig:Correlation}
\end{figure*}

	\begin{figure*}[htb]
	\centering
	\includegraphics[angle=0,scale=0.56]
{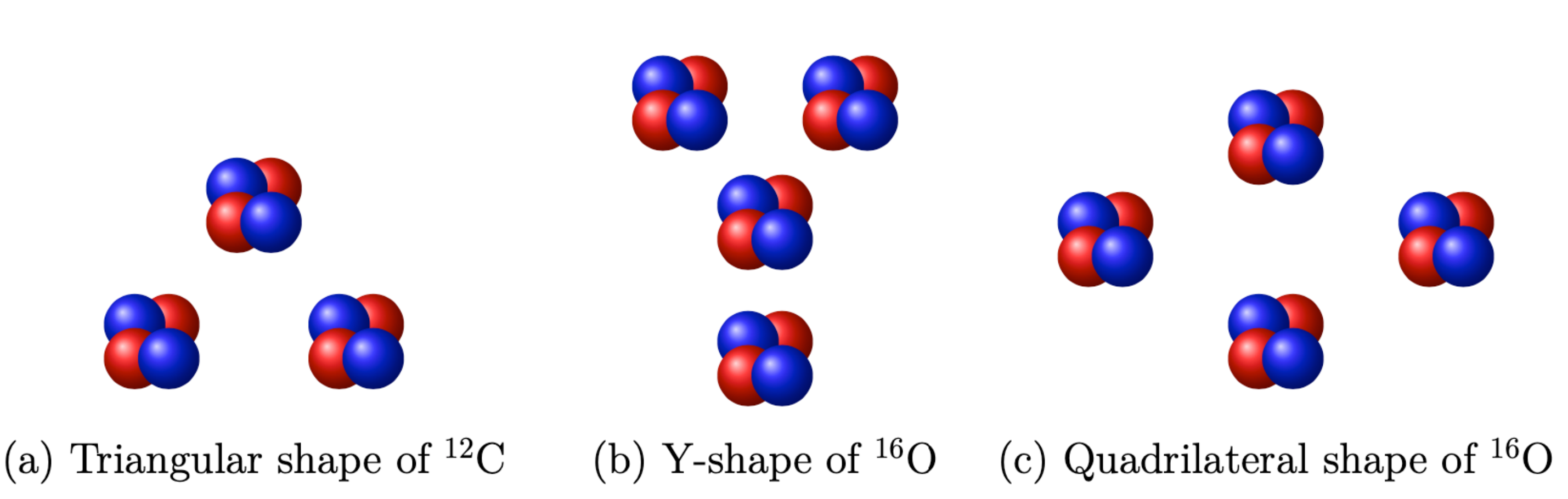}
	\caption {Schematic illustration of different shapes of clustered nuclei 
	in the transverse plane.
	}\label{fig:Shape}
\end{figure*}

Figure~\ref{fig:CorrelationAll} gives the correlation function of collision systems with different size. We can find a multi-peak structure on the away-side correlation in large systems such as $^{197}$Au + $^{197}$Au and $^{96}$Zr + $^{96}$Zr collisions in panel (a). The multi-peak structure is different from our previous work~\cite{ZHANG201176}, which was due to the mixed-event method to subtract background from the collective flow. In panel (b) the background is reconstructed by mixing events only with the requirement of similar second event plane, and the double-peak structure reappears in large collision systems as that in Ref.~\cite{ZHANG201176} as well as in many references, eg. \cite{WangPRL2,WangPRL3}. However, the triangular flow background contribution can result in the valley around $\Delta\phi$ $\sim$ $\pi$ of dihadron azimuthal correlation as Eq.~(\ref{eq2}) reveals for panel (b), which should be subtracted in order to get the reasonable away-side correlation information.

	Figure~\ref{fig:Correlation} shows the background-subtracted correlation functions in 0--2\% $^{12}$C + $^{12}$C collisions on panel (a) and (c), and $^{16}$O + $^{16}$O  collisions on panel (b) and (d) at $\sqrt{s_{NN}} = 6.37$ TeV. The collided nuclei are configured in the Woods-Saxon distribution and the $\alpha$-clustered structures, i.e. triangle for $^{12}$C and tetrahedron for $^{16}$O, respectively. We deducted both the second and the third order background in panel (a) and (b), while panel (c) and (d) just subtracted the second order background as a comparison. In the dihadron azimuthal correlation, the away-side multi-peak structure which appears in $^{197}$Au + $^{197}$Au collisions~\cite{aggarwalAzimuthalDihadronCorrelations2010} is absent in panel (a) and (b). 
In panel (c) and (d), although an obvious double-peak structure is shown in $^{16}$O + $^{16}$O collisions,  the multi-peak structure is still absent in the clustered $^{12}$C + $^{12}$C system, exhibiting a small system feature of light nuclei collisions as $^{10}$B + $^{10}$B  collision system in Fig.~\ref{fig:CorrelationAll}. Since the largest multiplicity comes with high triangularity in clustered nuclei collisions~\cite{broniowskiSignaturesClusteringLight2014}, especially for $^{12}$C with triangle-like arrangement, the influence of triangularity needs to be considered in this work. We adopt the  method with the third order background taking into account in the following calculations.

	All the dihadron azimuthal correlation functions exhibit noteworthy difference between the $\alpha$-clustered structure and the Woods-Saxon distribution on the away-side ($\Delta\phi \sim \pi$), where correlation signal is suppressed and broadened due to the violent interaction among the back jets and associated particles. The correlation yields in triangle $\alpha$-clustered $^{12}$C + $^{12}$C collisions is obviously higher, especially around $\Delta\phi \approx \pi$. In $^{16}$O + $^{16}$O collisions, the correlation function of tetrahedron $\alpha$-clustered $^{16}$O is slightly lower than the configuration with the Woods-Saxon distribution. By comparing results from the $\alpha$-clustered and ordinary nuclei in the two collision systems, it is seen that the dihadron azimuthal correlation results depend on the initial geometry structure of nuclei.

	To understand the effect from the $\alpha$-cluster structure on dihadron azimuthal correlations qualitatively, we can imagine an $\alpha$-clustered nucleus collides against another in the transverse plane via a geometric figure shown in Fig.~\ref{fig:Shape}.
	The high energy density areas around the $\alpha$ particles leads to incident hard scattering process and violent energy loss nearby. We can therefore consider $\alpha$ clusters as birthplace and barrier of jets. 
	Moreover, as mentioned in Refs.~\cite{broniowskiSignaturesClusteringLight2014, bozek__2014}, the orientation of clustered nuclei plays an important role in multiplicity. When the incoming nuclei keep the same orientation with the other and the coplane of $\alpha$ particles are parallel to the transverse plane, the $\alpha$ particles can collide directly, the clustered nuclei exhibit the largest overlapped region and the most wounded nucleons are produced.
	This situation can produce the highest multiplicity due to strong damage of nuclei, should be regarded as the main source of dihadron azimuthal correlations and studied primarily. Since the random orientation of the collision system, the most central collisions, 2\% centrality chosen in this work, can reflect the initial geometry properties from the nuclear structures. However for the lack of experimental data of the small collision system in this collision energy region, the exact configuration of parameters in the model for the particle production mechanism is an open question, which may cause model dependence.
	
	For the clustered $^{12}$C + $^{12}$C collision systems with high multiplicities, the 3$\alpha$ clusters construct a triangular shape  (Fig.~\ref{fig:Shape}(a)) in the transverse plane. 
	When a hard scattering process occurs at a vertex of the triangle, particles traveling outside straightly are more likely to hold high-$p_T$ property to be detected. The backside particles which pass through the other two vertexes loss energy furiously. Therefore, a sharper away-side peak appears.
	For clustered $^{16}$O + $^{16}$O collision systems, the two-dimensional projection of 4$\alpha$-particle condensates has a $Y$-like or quadrilateral shape in the transverse plane, as shown in Fig.~\ref{fig:Shape}(b) and (c). 
	If a trigger particle emerges inside the central vertex of the $Y$-shape structure, the amount of associated particles in all direction are expected to reduce due to the surrounding dense medium. The decreased yields can result in the suppression of correlation function on the near-side. Furthermore, for trigger particles which travel outwards from outer vertexes in both shapes, accompanying associated particles moving along the reverse direction will strike against the backside barriers. As a result, the away-side correlation peak around $\Delta\phi = \pi$ is restrained in the clustered condition.
	
	\begin{figure*}[thb]
	\centering
	\includegraphics[angle=0,scale=0.86]{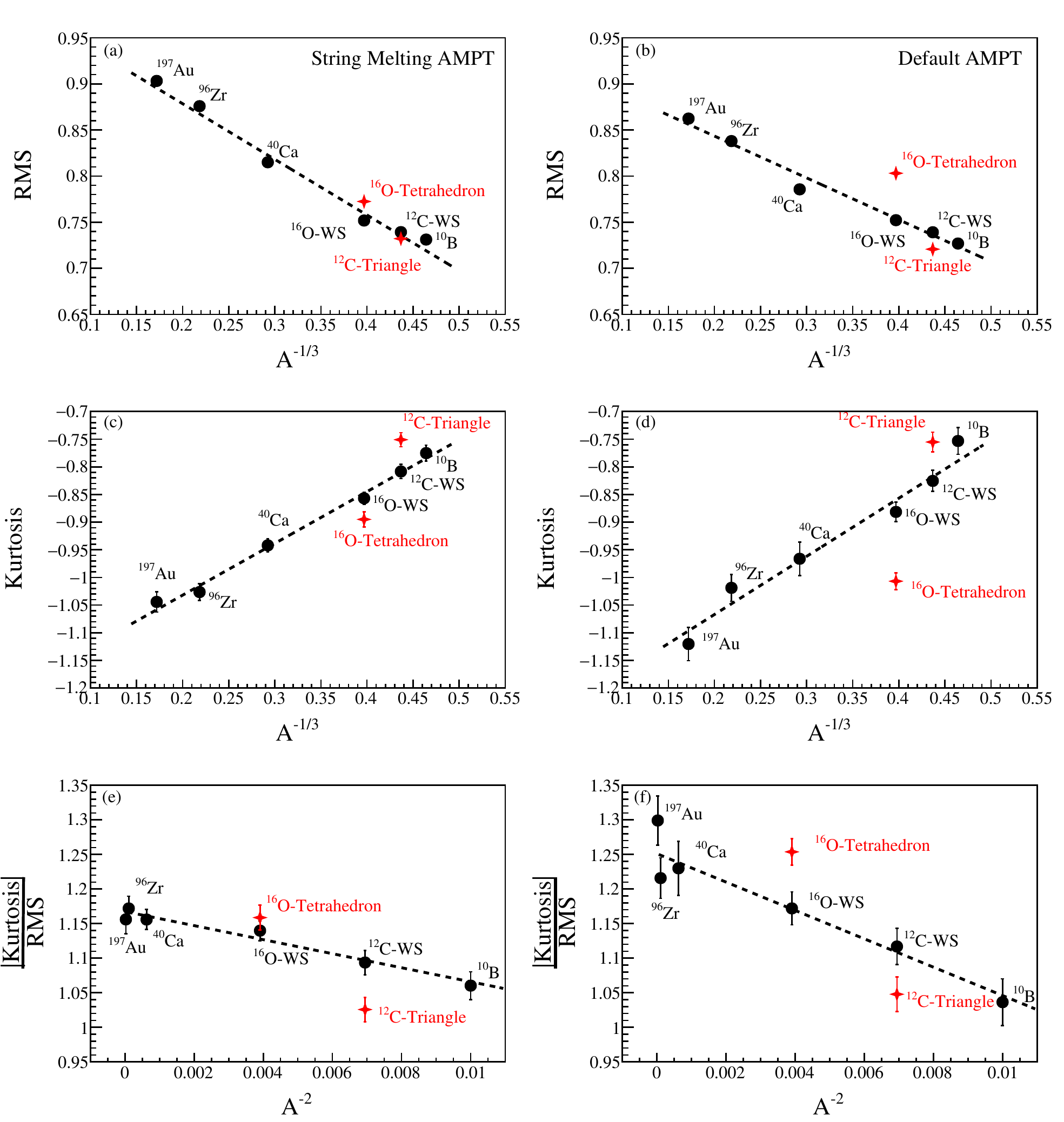}
		\caption{ System scan of away-side RMS width (a,b), Kurtosis (c,d) and $\frac{|\mathrm{Kurtosis}|}{\mathrm{Rms}}$  (e,f) of correlation functions in 0--2\% data at $\sqrt{s_{NN}} = 6.37$ TeV by the string melting (left panels) and default (right panels) AMPT model at $\sqrt{s_{NN}} = 6.37$ TeV. The statistical error bars of panel (a,b) are inside the marker size. The dashed lines in each panel are fitted as a function of  $\mathrm{k}\mathrm{A}^{-1/3} + \mathrm{c}$ with $k$ = -0.60 and $c$ = 1.00 (a) or $k$ = -0.45 and $c$ = 0.93 (b), 
	 $\mathrm{k}\mathrm{A}^{-1/3} + \mathrm{c}$ with $k$ = 0.94 and $c$ = -1.22 (c) or $k$ = 1.05 and $c$ = -1.28 (d), and 
 $\mathrm{k}\mathrm{A}^{-2} + \mathrm{c}$ with  $k$ = -1.02 and $c$ = 1.17 (e) or $k$ = -2.05  and $c$ = 1.25 (f). } \label{fig:SystemRms}
\end{figure*}

As discussed above, we observe the discrepancy on away-side dihadron azimuthal correlation for different collision systems, especially between clustered and uniform nuclei. Questions need to be asked, 1) how does the correlation reflect the violent degree  of interaction between the medium and high energetic particles in different size of system, and 2)  whether $\alpha$-clustering structure can be distinguished via dihadron azimuthal correlations in both $^{12}$C + $^{12}$C  and $^{16}$O + $^{16}$O collisions.  

	In order to pinpoint the discrepancy in the shape of correlation functions on the away-side,  the following cumulant extracted from away-side dihadron azimuthal correlation could characterize the distributions quantitatively. The root-mean-square (RMS) width \cite{ZHANG201176},  which describes the dispersion of the associated particles with respect to the direction of back jet, is defined as
	\begin{equation}
		\Delta\phi_{\mathrm{rms}} = \sqrt{\frac{\sum_\mathrm{away}(\Delta\phi-\Delta\phi_{\mathrm{m}})^{2}(dN/d\Delta\phi)}{\sum_\mathrm{away}(dN/d\Delta\phi)}},
	\end{equation}
	where $\Delta\phi_m$ is set as $\pi$, and the away-side region used in summation is from 1.5 to 2$\pi - 1.5$. Figure~\ref{fig:SystemRms}(a) shows the system size dependence of away-side RMS width as a function of $A^{-1/3}$, i.e. the reverse of the system size. Here $A$ means the system mass of the projectile (target is the same). In the collision systems with the  Woods-Saxon nucleon distribution, the RMS width is increasing smoothly with the increase of system size from $^{10}$B + $^{10}$B to $^{197}$Au + $^{197}$Au collisions, which presents the broadening distribution of associated particles in larger systems, indicating more violent interaction in larger size collision systems, which is consistent with our previous study~\cite{ZHANG201176}. This illustrates that the away-side width has a very good geometric origin: more larger medium size, more larger away-side width, which is consistent with the path-length effect as found before \cite{correlation5,ZHANG201176}.
	
	From figure~\ref{fig:SystemRms}(a), it is found that there are small differences of the RMS width between collision systems with the Woods-Saxon distribution and the $\alpha$-clustered structure by using the string melting AMPT model. The $\alpha$-clustered $^{12}$C + $^{12}$C collision has a smaller RMS width, while $\alpha$-clustered $^{16}$O + $^{16}$O collision system exhibits a larger value. The total away-side yields of two different kinds of $^{16}$O + $^{16}$O collisions are almost same, however, the $\alpha$-clustered system has a higher yield at the bottom and a lower yield at the peak. Hence the difference is strengthened. This can be attributed to the different structure in initial nucleon distributions as illustrated and discussed of Fig.~\ref{fig:Shape}. The set of red points which represent $\alpha$-clustered systems are close to the dashed line, and we can not get any conclusion via the similar value. However, the slight distinction of RMS width leads to our study for higher order moment.

	To further investigate the $\alpha$-clustering effect from the away-side correlation function, another useful observable refers to kurtosis~\cite{maAnisotropyFluctuationCorrelation2020}, which describes the tailedness of the distribution, is sensitive to correlation yields far from $\Delta\phi_m$ on the away-side. Kurtosis is defined as
	\begin{equation}
		\Delta\phi_{\mathrm{kurt}} =\dfrac{\sum_\mathrm{away}(\Delta\phi-\Delta\phi_{\mathrm{m}})^{4}(dN/d\Delta\phi)}{ \Delta\phi_{\mathrm{rms}}^4 \cdot \sum_\mathrm{away}(dN/d\Delta\phi)} - 3,
	\end{equation}
	in which the summation takes the same computing steps as RMS width. This definition sets the kurtosis of normal distribution equal to zero. We understand that kurtosis is sensitive to the value at the tail of the peak. Nevertheless, when the away-side correlation functions are almost same at the  bottom, kurtosis can still tell the difference in magnitude around $\Delta\phi \sim \pi$. The results of kurtosis in different systems are displayed in Fig.~\ref{fig:SystemRms}(c) as a function of $A^{-1/3}$. All negative values of kurtosis for different collision systems illustrate that the away-side dihadron distribution widths are all wider than the Gaussian distribution. And a clear trend of decreasing kurtosis with increasing system size in uniform heavy-ion collision systems demonstrates that the flattening correlation peak on the away-side, especially the augmenting associated particles perpendicular to the direction of jet.

	It is noticeable that the disparity of kurtosis between the clustered and uniform nuclei is more remarkable than RMS width.
	For the $\alpha$-clustered $^{12}$C collisions, it has the same yield at the bottom and a higher yield at the peak. The RMS difference is neutralized due to the normalization, whereas the kurtosis difference is strengthened. 
	The results mean that the kurtosis of away-side dihadron correlation could be more capable of sorting the clustered nuclei. 
	
	\begin{figure*}[htb]
	\includegraphics[angle=0,scale=0.8]{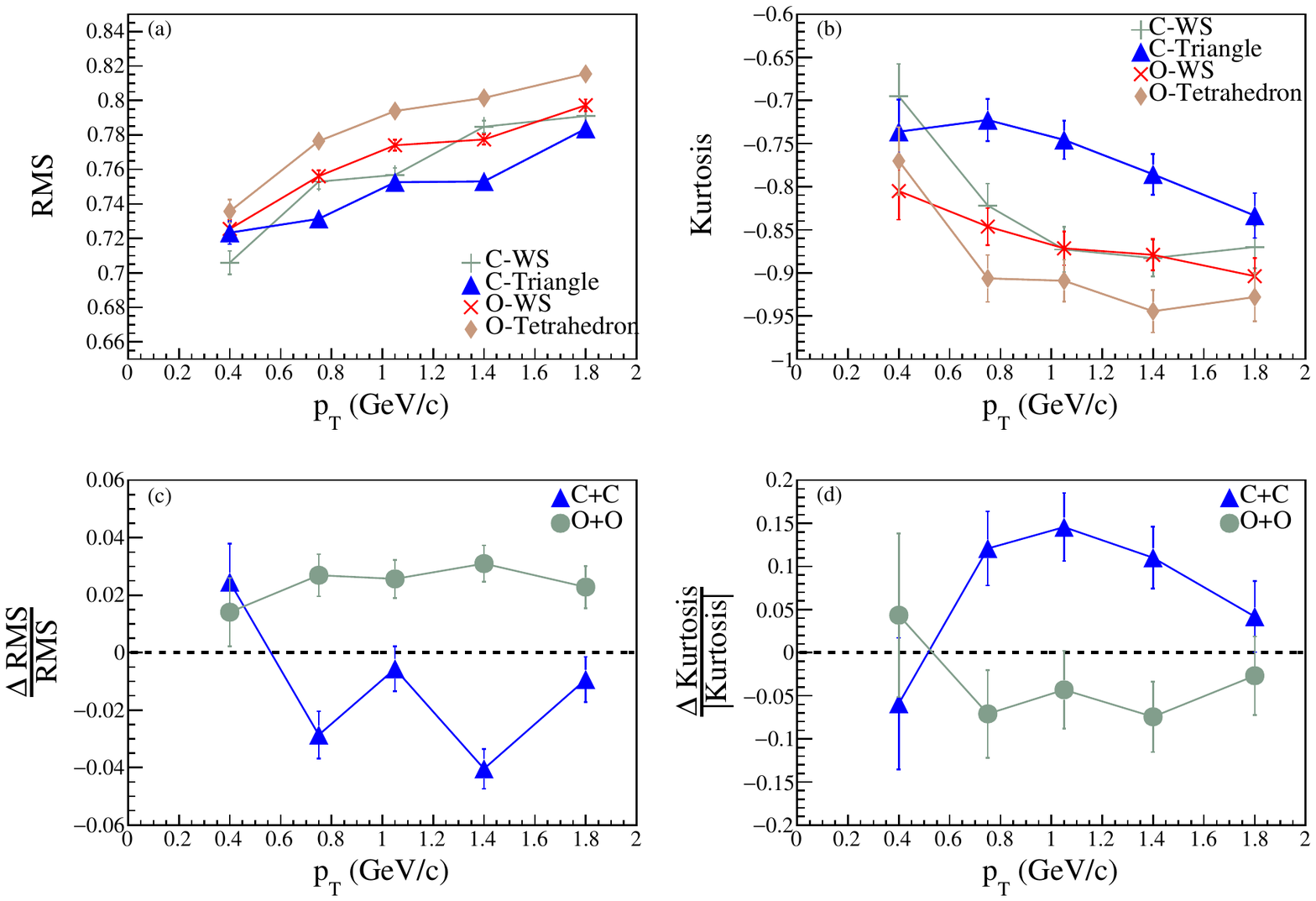}
		\caption{  RMS width (a), kurtosis (b), relative difference of RMS width (c) and kurtosis (d) versus $p_T^{\mathrm{asso}}$ in $^{12}$C + $^{12}$C  and $^{16}$O + $^{16}$O collision systems  in 0--2\% data at $\sqrt{s_{NN}} = 6.37$ TeV.} \label{fig:PtDependence}
\end{figure*}

	In addition, we study the ratio of $|$Kurtosis$|$ over RMS versus $A^{-2}$ as shown in Fig.~\ref{fig:SystemRms}(e) . Similar to RMS and kurtosis, the values of the Wood-Saxon configurations are matched in a smooth spline which can be fitted with a function of $A^{-2}$. 
	This  ratio from the clustered nuclear collisions are disengaged from the spline and can be also served as a probe.
	
	As discussed above the lack of the experimental data for small collision systems in this collision energy  region, it is informative to compare  partonic interaction dominated process and hadron interaction dominated process in the model framework.
	It is also essential to investigate how the partonic interaction affects the transforming of the initial properties into the final state. Here we employed the default AMPT model, which turns off the partonic interaction in some extents, to investigate the effect of the partonic interaction on the dihadron azimuthal correlation.
Right panels (b, d, f) of Fig.~\ref{fig:SystemRms} present the default AMPT results in the system scan with the same configuration as introduced before. There are two facts we would mention.
The first point is that we observed the similar  power-law scalings of the RMS width, Kurtosis as well as $|Kurtosis|/RMS$ between two AMPT versions for the nuclei with the Woods-Saxon  distribution. But the slopes of scaling relation look different, especially for the RMS. Larger slope for the melting case indicates that partonic interaction becomes more important in larger system. The second point is that we observed a strong deviation between red points which represents the $\alpha$-clustered nuclei and dashed line fitted by the Woods-Saxon systems. 

Via comparison of the default and the melting AMPT model, although the value of observables changed a bit, the differences between the Woods-Saxon and the $\alpha$-cluster cases exhibit a more obvious signal especially for $\alpha$-clustered $^{16}$O + $^{16}$O system, i.e. it leads to better distinguish the Woods-Saxon case from the $\alpha$-clustering case without partonic interactions. This distinction between the default and the melting version demonstrates that the behavior of dihadron azimuthal correlation is sensitive to whether  the system undergoes a partonic interaction process or not ~\cite{ZHANG201176}. 
The reason could be related to the particle production mechanism and the transport properties in the fireball. The high momentum particles in jets were produced at early stage, simulated by HIJING model in this calculation, which will carry the information of the initial geometry distributions. 
At the evolution stage, the geometrical asymmetry will be transferred to momentum space, where the initial properties will be lessened according to the violent extent of interaction. 
Moreover, partonic process of the string melting AMPT produce higher yield of particles in low $p_{\mathrm{T}}$ region  as well as  stronger elliptic flow which participate in partonic scatterings, then it  leads to  a larger background and blurs the initial signals \cite{linMultiphaseTransportModel2005, maDihadronAzimuthalCorrelation2006}.
As shown in right panels of Fig.~\ref{fig:SystemRms}, the results from $\alpha$-clustering case are obvious deviated from the Woods-Saxon case in the default AMPT, which mainly reflects the initial properties from the early hard process. The relative difference between the two cases becomes minor with the evolution in partonic cascade where the high momentum particles as well as the soft particles in the medium will exchange energy each other and the impact of initial shape  
tends isotropic gradually~\cite{EPJA.54.161-SZhang2018}.
In a word, the initial geometry structure effect becomes slightly faded after the system undergoes partonic interaction but still remains in the final state.

	The $\alpha$-clustering effects on dihadron azimuthal correlation preserve a wide range of the transverse momentum. Here, we select the trigger particle with 2 $<p_T<$ 6 GeV/$c$, and change the momentum range for the associated particles continuously. The background of dihadron correlation is evaluated with event plane method. In Fig.~\ref{fig:PtDependence}(a) the RMS width increases  with transverse momentum of associated particles in all collision systems. From Fig.~\ref{fig:PtDependence}(b), the kurtosis is expected to decline gradually with the increasing of transverse momentum of associated particles and tends to reach a constant in small collision systems.  The relative difference of them are also displayed in Fig.~\ref{fig:PtDependence}(c) and (d). Although the paired particles are selected with different momentum, the magnitude relation among common collision systems does not change in general. The larger collision system has lager RMS width and lower kurtosis. Clustered and uniform distribution of $^{12}$C and $^{16}$O can also be identified just by the value of kurtosis as well as RMS  width. Indeed, we demonstrate that the range of $p_T^{\mathrm{asso}}$ is irrelevant to the differences of different nuclei in the shape of away-side correlation functions.

Since the traversing path length of particles with different direction is expected to depend on the event centrality, the centrality dependence of dihadron azimuthal correlation is also checked for centrality bins of 0--2\%, 2--5\% and 5--10\%. It is found that the relative differences of the RMS width and kurtosis reach the largest value in the most central collisions in general. This represents the discrepancy shown in the clustered nuclear systems can benefit from longer path length.
Assuming that all collisions are completely central with b = 0 fm, the largest collision region is created, the initial distribution of nucleons will take the maximum preservation in the fireball, and the jet quenching or parton energy loss will reach the maximum extent. Consequently, the most central collisions are preferred in such kind of study for the experiments.
	
\section{\label{sec:Summary}Summary}
	This paper presents a collision system scan involving $\alpha$-clustered $^{12}$C and $^{16}$O by using a multiphase transport model for  central collisions from 
$^{10}$B + $^{10}$B to $^{197}$Au + $^{197}$Au 	
	at $\sqrt{s_{NN}} = 6.37$ TeV.  Away-side broadening of dihadron  azimuthal correlation is observed via the second and third order background subtracted from the raw signals, and the related RMS width and Kurtosis parameters are found to follow the $A^{-1/3}$ law of the system size if the nucleus has the normal Woods-Saxon nucleon distributions. Furthermore, 
	the effect of $\alpha$ clustering structure on dihadron azimuthal correlation is visible by the deviation from the above mentioned $A^{-1/3}$ law, which provides a potential observable of $\alpha$-cluster structure for experimental analysis. 
	This system size and configuration dependences are related to the path length along which the energetic particles pass through the medium created in the collision. The evolution of the dense medium makes the spatial anisotropy transform to the momentum space, especially induce the broadening of away-side correlation structure.
	
	In addition, we adopt the default AMPT version to check if partonic interaction, which also reflects the partonic energy loss mechanism in the medium, makes difference. By comparing results with or without parton cascade, it is revealed that the partonic interaction lessens the impact of initial structure but is less of a crucial factor on the distinction between the Woods-Saxon and the $\alpha$-cluster cases. Meanwhile, the associated particle transverse momentum dependence of away-side correlation functions is investigated by computing root-mean-square width and kurtosis. 
	The manifestation of $\alpha$-clustered nuclei with varying transverse momentum shows a significant divergence with respect to uniform nuclei. The independence with transverse momentum on the relation among different systems strengthens the applicability of dihadron azimuthal correlation to probe the $\alpha$-clustering structure in relativistic heavy-ion collisions.  Centrality dependence of dihadron azimuthal correlation is also studied basically, and 
it is found that the relative differences reach the largest value in the most central collisions in general. 

In conclusion, 
a system scan project in central collisions is expected to reveal the properties of the evolution of the phase-space from initial to final state in the collisions as well as to distinguish the exotic nuclear structures like $\alpha$-clustering phenomenon in light nuclei.

\vspace{.5cm}
	
{\bf Declaration of competing interest}: The authors declare that they have no known competing financial interests or personal relationships that could have appeared to influence the work reported in this paper.

\begin{acknowledgments}
This work was supported in part by Guangdong Major Project of Basic and Applied Basic Research No. 2020B0301030008, the National Natural Science Foundation of China under contract Nos.11890710, 11890714, 12147101, 11875066, 11925502, 11935001, 11961141003 and12061141008,  National Key R\&D Program of China under Grant No. 2018YFE0104600, and the Strategic Priority Research Program of CAS under Grant No. XDB34000000.
\end{acknowledgments}

\end{CJK*}

\bibliography{Ref}

\end{document}